# Full-Scale Indexing and Semantic Annotation of CT Imaging: Boosting FAIRness

Hannes Ulrich[1], Robin Hendel[2], Santiago Pazmino[1], Björn Bergh[1] & Björn Schreiweis[1]

1. Institute for Medical Informatics and Statistics, Kiel University and University Hospital Schleswig-Holstein, Campus Kiel, Kiel and Lübeck, Schleswig-Holstein, Germany
2. University Hospital Würzburg, Würzburg, Germany

## Abstract

- **Background:** The integration of artificial intelligence into medicine has led to significant advances, particularly in diagnostics and treatment planning. However, the reliability of AI models is highly dependent on the quality of the training data, especially in medical imaging, where varying patient data and evolving medical knowledge pose a challenge to the accuracy and generalizability of given datasets.
- **Results:** The proposed approach focuses on the integration and enhancement of clinical computed tomography (CT) image series for better findability, accessibility, interoperability, and reusability. Through an automated indexing process, CT image series are semantically enhanced using the TotalSegmentator framework for segmentation and resulting SNOMED CT annotations. The metadata is standardized with HL7 FHIR resources to enable efficient data recognition and data exchange between research projects.
- **Conclusions:** The study successfully integrates a robust process within the UKSH MeDIC, leading to the semantic enrichment of over 230,000 CT image series and over 8 million SNOMED CT annotations. The standardized representation using HL7 FHIR resources improves discoverability and facilitates interoperability, providing a foundation for the FAIRness of medical imaging data. However, developing automated annotation methods that can keep pace with growing clinical datasets remains a challenge to ensure continued progress in large-scale integration and indexing of medical imaging for advanced healthcare AI applications.

## Background

In recent years, integrating artificial intelligence (AI) into medicine has made promising progress and revolutionized diagnostics, treatment planning, and healthcare management [1]. However, the effectiveness of AI models depends heavily on the quality and reliability of the data sets used to train them. Reliable and real-world clinical data play a crucial role in ensuring the accuracy and generalizability of AI in medicine. Patient data can vary significantly due to factors such as age, gender, genetics, and lifestyle. In addition, healthcare professionals often deal with rare or novel cases that are not well represented in standard datasets [2]. Therefore, reliable datasets must express the clinical reality and be continuously updated so that models can adapt to new medical challenges. This adaptability is crucial for the robust performance of AI systems in the face of evolving medical knowledge and the discovery of previously unknown diseases. Most medical AI models are based on radiological imaging, an important diagnostic tool. However, currently, existing datasets are rarely updated, so approaches are needed to provide up-to-date and customized datasets that reflect real-world clinical environments.

The IMPETUS junior research group is focusing on precisely these challenges. The main objective of the IMPETUS group is to upgrade the Medical Data Integration Center (MeDIC) [3], which was set up at the UKSH as part of the HiGHmed consortium [4], to enable the integration and reuse of all multimedia objects and reports, regardless of format, storage or presentation in a standardized environment. In a prior study, we integrated the productive picture archiving and communication system (PACS) of University Hospital Schleswig-Holstein, the second-largest university hospital in Germany, and established an automatic integration process to retrieve new imaging series and metadata on a daily basis [5]. By February 2024, over 33 million clinical imaging series had been successfully integrated. Unfortunately, the metadata to identify the corresponding data for incoming research requests is incomplete or missing. The metadata is not routinely curated for the purpose of findability, and radiological reports are often semi-structured or focused on relevant pathologies (without listing all visible structures). However, for research and for the creation of training data, an annotation and indexing process can improve the discoverability of relevant anatomical data. In this study, we are focusing on a subset of the imaging data, the computer tomography (CT) imaging series, and present our approach to increase the findability, accessibility, interoperability, and reusability (FAIRness) of the clinical imaging series [6]. The goal is the conceptualization and establishment of a robust indexing process to improve the metadata of the integrated CT imaging series. The incoming series and the corresponding metadata should be semantically enhanced to provide more granular search criteria to discover relevant imaging series within the growing dataset.

## Implementation

The integration and indexing process must seamlessly integrate with the established MeDIC architecture, ensuring compatibility and continuity. The indexing process should exhibit robustness, particularly in handling diverse CT imaging protocols. Time efficiency is paramount, with computing requirements below the daily incoming volume and fully scalable to allow the processing of legacy data in parallel. Results must conform to standardized formats, facilitating adherence to national and international data-sharing initiatives and thereby enhancing accessibility and reusability of the data for researchers.

### Image Retrieval and Semantic Enhancement

The process is triggered by the daily image series integration [5]. The implemented process is embeddable in the current MeDIC architecture, as seen in Figure 1. The daily CTs are indexed fully automatically in a downstream process flow. For each incoming CT series, an indexing task is created and transferred to a Kafka topic. The task will be received by an indexing service instance, loading the series from the PACS, segmenting it, and analyzing it. The indexing service is based on the TotalSegmentator introduced by Wasserthal et al. [7]. Their framework segments 104-124 anatomical structures in CT datasets, depending on the version used. The implemented process is embeddable in the current MeDIC architecture, as seen in Figure 1. The service uses the TotalSegmentator full-body task. The result of the segmentation process is a single statistics file since the actual segmentation masks are discarded, respectively, or not saved during the process. This statistic file consists of all identified anatomical structures, their volumes in mm$^3$, and the grayscale intensity. To provide the results semantically enhanced, a mapping from TotalSegmentator labels to SNOMED CT and Radlex is provided. The mapping is established using the methodology described by ISO/TR 12300:2014 "Health informatics - Principles of mapping between terminological systems," [8] and evaluated by a radiological specialist.

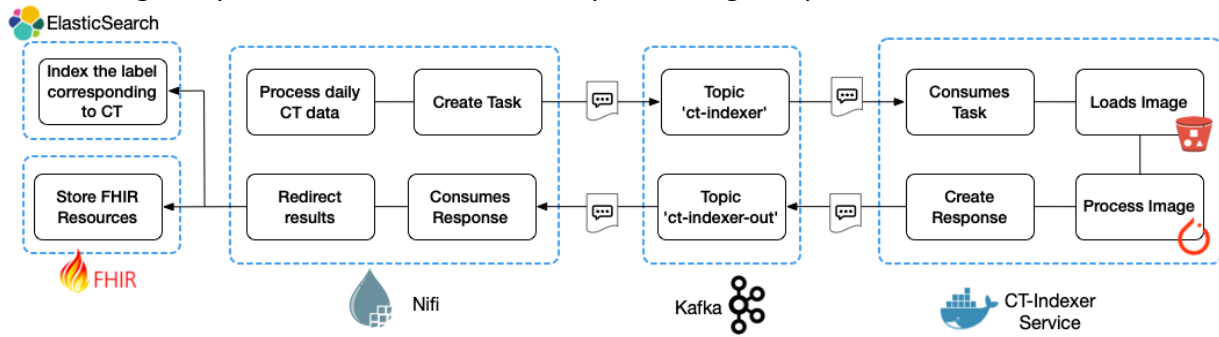

*Figure 1 Technical overview of the implemented indexing process within the UKSH MeDIC. The process is started with Apache Nifi creating a task based on the incoming data. The task is forwarded to the indexing services via Kafka. After the successful indexing the results are sent back via Kafka and forwarded to the designated destination.*

## Standardized representation using HL7 FHIR

For the MeDIC internal indexing process, the annotation file is enhanced with the established mapping and stored within the central ElasticSearch of data lake to enable fast findability for the downstream processing. To enable a more advanced and interoperable search, the annotation file is transformed into a standardized representation using HL7 FHIR resources [9] containing patient information, imaging metadata, and the corresponding SNOMED CT annotations, shown in Figure 2. The software versions of the indexing service and relevant libraries used were modeled in the Device, more precisely in the included *DeviceVersion*. Annotations and versions are linked via a Provenance resource so that future version changes can be tracked and comprehensible. The resource definitions are formalized in the FHIR Shorthand language [10] and compiled into a set of FHIR profiles using sushi [11] and are available on GitHub [12].

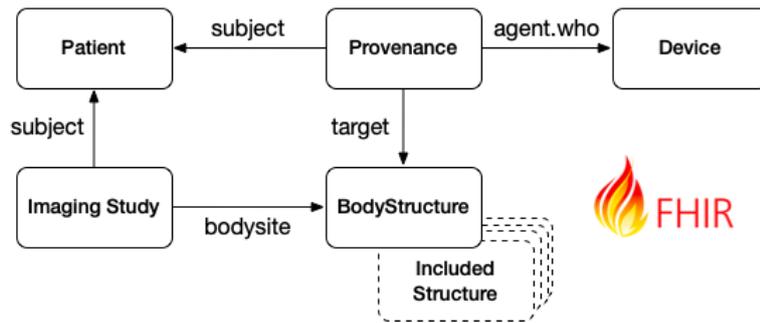

*Figure 2 All associated resources and the used references for CT-Indexer profiles. The resources Patient and ImagingStudy include general information to identify the patient and the corresponding imaging series. The annotations are represented as a Body structure resource, including the assigned SNOMED CT codes. The Device and Prov Resource contain process information and ensure retrospective traceability of the annotations.*

## Results

We successfully included the process in the established imaging integration flow and enabled automatic indexing for all CT imaging series within the USKH MeDIC. Until March 1, 2024, 230,000 CT imaging series are semantically enhanced using the presented approach. In total, more than 8,528,000 SNOMED CT annotations were added to the image series. The indexing service runs in eight instances in parallel, resulting in an average throughput of 200 imaging series per hour and over 155 days of GPU computing time. We established a priority-based processing sequence to index the imaging data from the previous day primarily. While the daily data are processed, the legacy data are continuously retrieved from the repository in chronological order and indexed until the entire PACS system is fully semantically enriched. The standardized representation using HL7 FHIR R5 yields a maximum of five different FHIR resources per CT imaging series. The less dynamic resources, e.g., *Patient* and Device, are defined as *conditional creates* to reduce redundancy within the FHIR export and downstream repository. The 230,000 indexed series results in ca. 750,600 unique FHIR resources that are stored on a separate server to enable effective data discovery. The indexing service is open-source and freely available on GitHub [13].

## Discussion

In the field of medical imaging, the ability to efficiently navigate and discover vast datasets is paramount. This necessitates sophisticated methodologies that enable fine granular search within the exponentially expanding datasets. Moreover, the standardization and semantic enhancement of imaging series representation are crucial for fostering interoperability and facilitating seamless data exchange among diverse research projects. Currently, most of the imaging series are hardly semantically annotated since the task is time-consuming and resource-intensive. However, radiological imaging continues to be one of the most clinically useful tools in various clinical fields and research projects, such as the European Health Data Space [14], which emphasizes the importance of engaging in standardized exchange formats to enable their cross-institutional findability and reuse. Our approach aims to strengthen the four FAIR principles: (F) via the continuous integration of the PACS, the imaging series are retrievable; (A) due to an established application process [12, 13] the data can be requested for medical research; (I) and (R) The use of FHIR as a standardized exchange format in combination with automated SNOMED CT annotations strengthens interoperability and reusability [17].

Traditionally, automatic image annotation has been a cornerstone in downstream data analysis processes. With the emergence of artificial intelligence and the adaption of various classification and segmentation networks, the development and exploration of traditional automatic annotation methods have slowed down, as these approaches often struggle with work effort and maintaining accuracy in growing clinical datasets.

The use of machine-learning methods for semantic annotation is logical due to the current research momentum – the use of segmentation networks for semantic annotation seems unusual at first thought, but the frameworks are remarkably robust on clinical data. The integrated imaging series are of wide variance due to the use of different CT devices and recording protocols over the decades. It was, therefore, necessary to apply a steady and robust method: the TotalSegmentator is a state-of-the-art framework that achieves high accuracy (Dice coefficient of 0.943) and outperforms other available segmentation frameworks in terms of label variety. Additionally, it has broad applicability and robust performance, especially in clinical settings. The framework shows a robust performance overall; only the processing of large full-body CTs is unstable due to a bug in the underlying library. The proposed service uses the functionality to index the imaging series in the first place but also has a segmentation mode to provide the segmentation mask to researchers as a MeDIC service if requested. The implemented service is just wrapped around the TotalSegmentator, so in the future, a change of the used network can be easily established. In the current research momentum, new frameworks with more labels [18] or a more robust performance are very likely. The presented approach is focused on CT as the processed modality. But future services shall index independently of modality [19] and thus enable consistent indexing across the entire PACS would be conceivable.

## Conclusions

The integration of artificial intelligence into medicine has made remarkable progress and revolutionized diagnostics and treatment planning. However, the effectiveness of the current models depends on a continuous flow of representative training data from clinical routine in order to keep the models relevant. Our efforts to enhance the metadata semantically for CT image series within UKSH MeDIC are a crucial step towards improving findability, accessibility, interoperability, and reusability – FAIRness – in medical imaging. Significant progress has been made through the semantic preparation of more than 230,000 CT image series, enabling a better discovery and resulting in a standardized representation using HL7 FHIR resources. The development of robust methods for automatic annotation remains paramount, especially in the context of growing clinical datasets, to ensure the continuing progress of large-scale integration and indexing of medical imaging.

## Availability and requirements

**Project name:** IMPETUS CT Indexer
**Project home page:** https://github.com/IMIS-MIKI/impetus-ct-indexer
**Operating system(s):** Platform independent
**Programming language:** Python
**Other requirements:** Python 3.10, GPU is recommended
**License:** Apache-2.0
**Any restrictions to use by non-academics:** none

# List of abbreviations

HL7:   Health Level 7
FHIR:  Fast Health Interoperability Resources
AI:    Artificial Intelligence
MeDIC: Medical Data Integration Center
PACS:  Picture Archiving and Communication System
CT:    Computer Tomography

# Acknowledgments

Not applicable.

# Funding

This research was funded by the German Federal Ministry of Education and Research, grant number 01ZZ2011 and 01KX2121. The funding agency had no role in study design, data collection, data analysis, results interpretation, or in writing the manuscript.

# Availability of data and materials

Not applicable.

# Author information

## Authors and Affiliations

Institute for Medical Informatics and Statistics, Kiel University and University Hospital Schleswig-Holstein, Campus Kiel, Kiel and Lübeck, Schleswig-Holstein, Germany

Hannes Ulrich, Santiago Pazmino, Björn Bergh & Björn Schreiweis

University Hospital Würzburg, Würzburg, Germany
Robin Hendel

## Contributions

HU developed the indexing concept and implementation. HU and RH established and evaluated the semantic mapping. SP implemented the scalability of the service. BB and BS contributed conceptually and conducted review and editing. All authors contributed to the writing of the manuscript. All authors read and approved the final manuscript.

Corresponding author

Hannes Ulrich (hannes.ulrich@uksh.de)